\documentclass{PoS}

\usepackage{epsfig,mathbbol}

\PoS{PoS(LAT2005)305}

\title{Laplacian modes for calorons and as a filter 
\newline
\vspace*{-4cm}
\rightline{\large\rm HU-EP-05/57}
\vspace{1.6cm}
}

\ShortTitle{Laplacian modes for calorons and as a filter }

\author{\speaker{Falk Bruckmann}\\
        Leiden University\\
        E-mail: \email{bruckmann@lorentz.leidenuniv.nl}}

\author{Ernst-Michael Ilgenfritz\\
        Humboldt University, Berlin\\
        E-mail: \email{ilgenfri@physik.hu-berlin.de}}

\abstract{We compute low-lying eigenmodes of the gauge covariant
  Laplace operator on the lattice at finite temperature. For classical
  configurations we show how the lowest mode localizes the monopole
  constituents inside calorons and that it hops upon changing the
  boundary conditions. The latter effect we observe for thermalized
  backgrounds, too, analogously to what is known for fermion zero
  modes.

 We propose a new filter for equilibrium configurations which provides
 link variables as a truncated sum involving the Laplacian modes. This
 method not only reproduces classical structures, but also preserves
 the confining potential, even when only a few modes are used.}

\FullConference{XXIIIrd International Symposium on Lattice Field Theory\\
                 25-30 July 2005\\
                 Trinity College, Dublin, Ireland}

\begin{document}

\section{Introduction}

We study low-lying eigenmodes of the gauge
covariant Laplace operator 
\begin{eqnarray}
-\triangle\phi_n(x)=\lambda_n\phi_n(x)\,,
\qquad
\triangle^{ab}_{xy}=\sum_{\mu=1}^4\left[
U_\mu^{ab}(x)\delta_{x+\hat{\mu},y}+
U_\mu^{\dagger ab}(y)\delta_{x-\hat{\mu},y}-
2\delta_{ab}\delta_{xy}\right]
\label{eqn_1}
\end{eqnarray}
with $U_\mu(x)$ a given lattice configuration in
the fundamental (or adjoint) 
representation of $SU(2)$. We will use these modes as
an analyzing tool, like it has been done with fermionic (near) zero modes. The
latter owe their existence to the index theorem, and in smooth cases
they are localised to the nontrivial background. Laplacian modes are
computationally cheaper because they are space-time scalars
and do not suffer from chirality problems nor doubler problems.
They seem not to be 
directly related to topology, but it has been observed that they are
sensitive to the location of instantons \cite{bruckmann:01a,deforcrand:01a}. 
The scaling
properties of their localization have been investigated recently
\cite{greensite:05}.

Here we will present two ideas concerning Laplacian modes
\cite{bruckmann:05b}. 
The first one is to study what can be learnt from
the profile of the modulus of the lowest mode 
(the `ground state probability density')
$|\phi_0(x)|^2$, in particular with changing boundary conditions. The second
idea is to introduce a novel low-pass filter. It concerns the reconstruction of
the gauge background from a few Laplacian modes. The purpose of both
methods is to identify the underlying infrared degrees of freedom
in the gauge field
that are responsible for features like confinement.

\section{Profile of the ground state density}

\subsection{Caloron backgrounds}

\begin{figure}[b]
\includegraphics[width=0.45\linewidth]
  {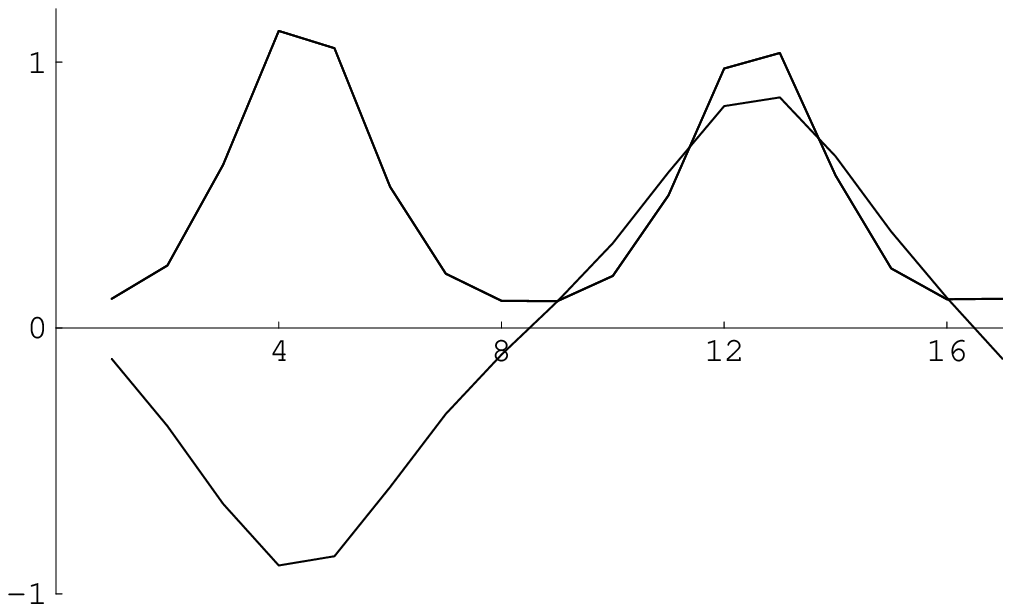}
\includegraphics[width=0.45\linewidth]
{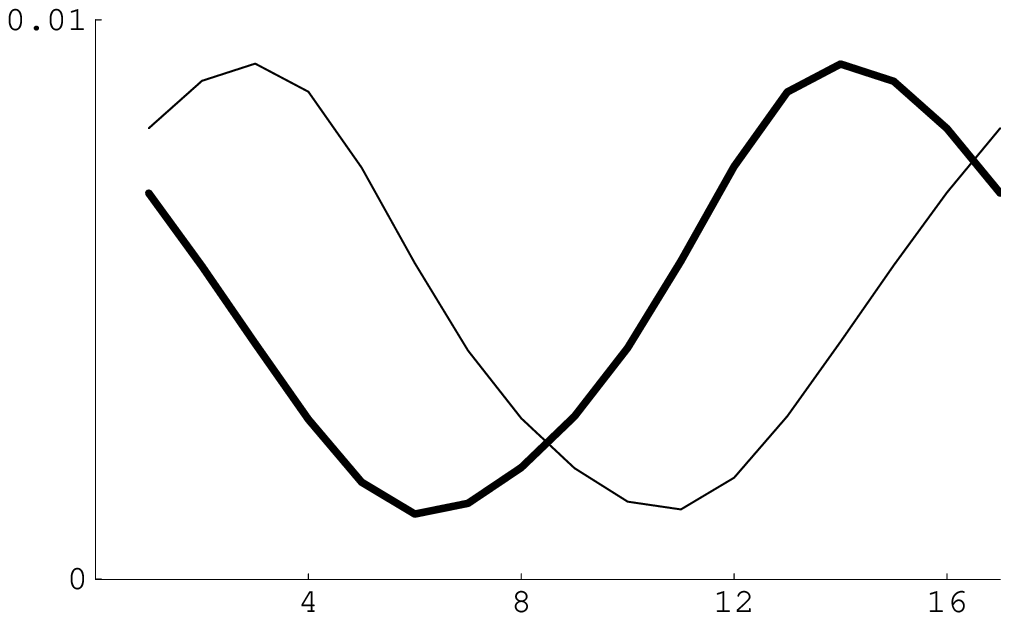}
\caption{A caloron with its constituent
monopoles. Left: the action and topological
  density (not distinguishable, both multiplied by a factor 400) and
  the Polyakov loop. Right: the modulus of the lowest Laplacian mode
  with periodic (boldface line) and antiperiodic (thin line) boundary conditions.}
\label{fig_1}
\end{figure}

As a testing ground for the Laplacian modes we first investigate 
a caloron of maximally nontrivial holonomy 
with its monopole constituents \cite{kraan:98a,lee:98b} 
put on a $16^3\cdot 4$ lattice\footnote{
This is done by calculating links from 
the continuum gauge field, followed by a few steps of cooling.}.
 Fig.\ \ref{fig_1} (left) shows the action 
density along the line connecting these monopoles, which are
clearly visible as two selfdual (and almost perfectly static) lumps.
As another signal the Polyakov loop goes through $\Eins_2$ and
$-\Eins_2$ -- which amounts to a local symmetry restoration -- at the
monopole cores.

The ground state of the Laplacian in this background is shown in Fig.\
\ref{fig_1} (right). One can see that the presence of monopoles
is reflected by a maximum resp.\ a minimum in the profile of the
lowest mode (although shifted by up to 2 lattice spacings).
In addition to the mode periodic in Euclidean time we have depicted
the antiperiodic one. In this mode the role of the monopoles are
interchanged, which can be understood from a symmetry of the caloron.
Hence the lowest-lying Laplacian mode `hops' as the result of changing the
boundary conditions. It behaves
similar to the caloron fermion zero mode \cite{garciaperez:99c}
in the context of which complex boundary conditions were introduced first.

\subsection{Thermalized backgrounds}

\begin{figure}[b]
\includegraphics[width=0.32\linewidth]
{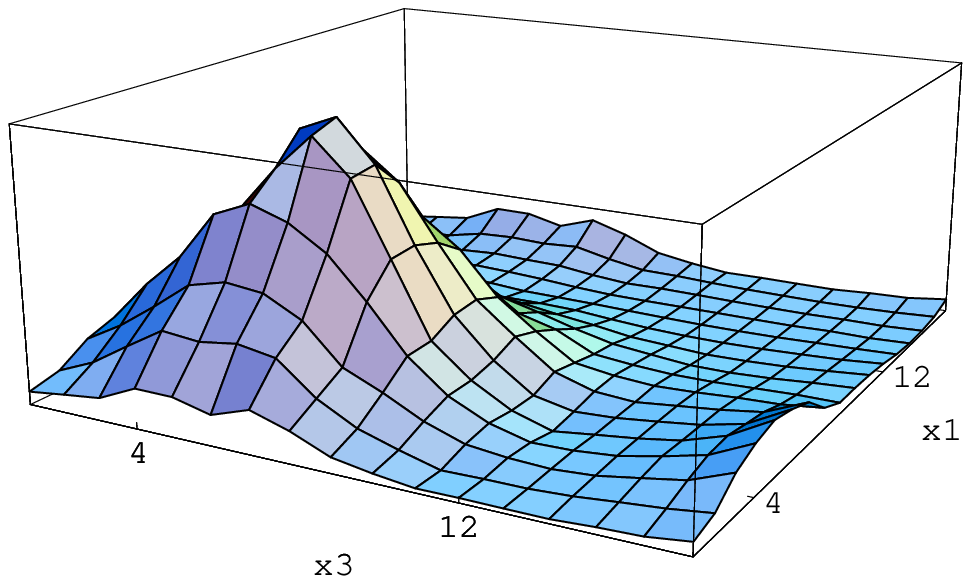}
\includegraphics[width=0.32\linewidth]
{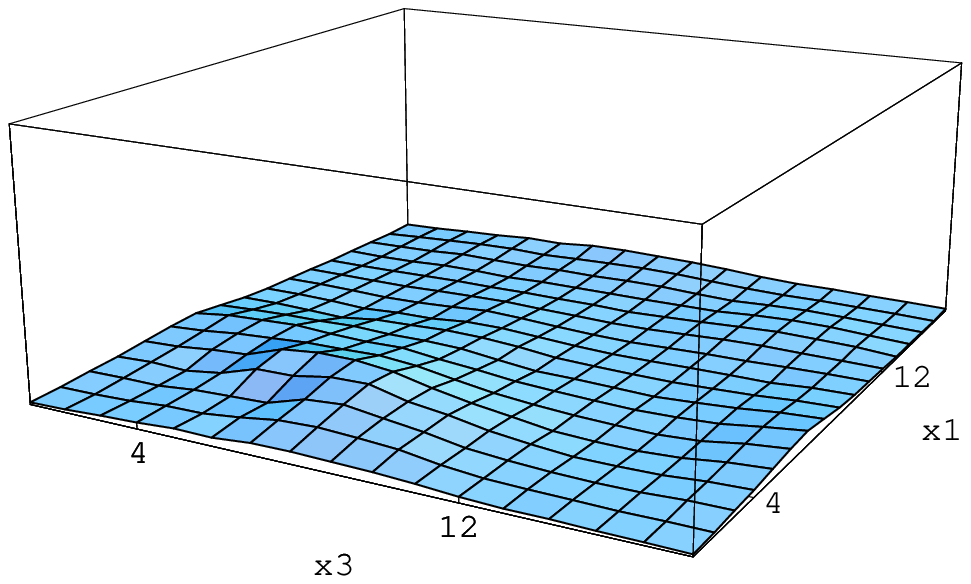}
\includegraphics[width=0.32\linewidth]
{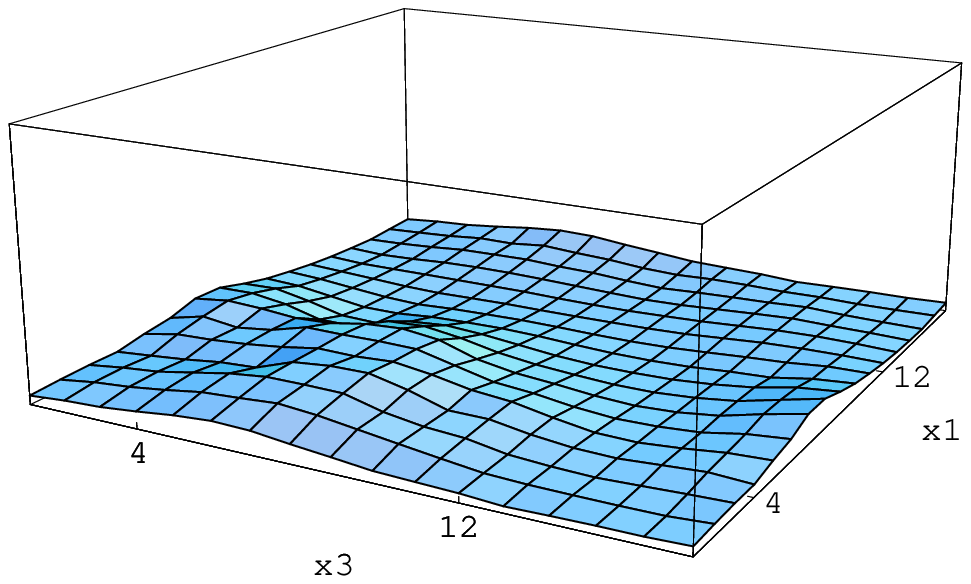}\\
\includegraphics[width=0.32\linewidth]
{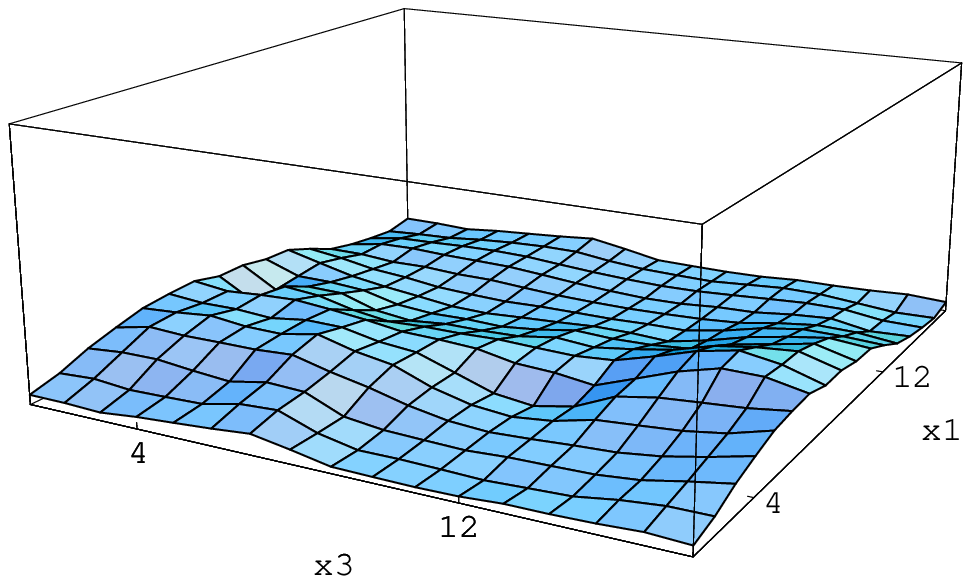}
\includegraphics[width=0.32\linewidth]
{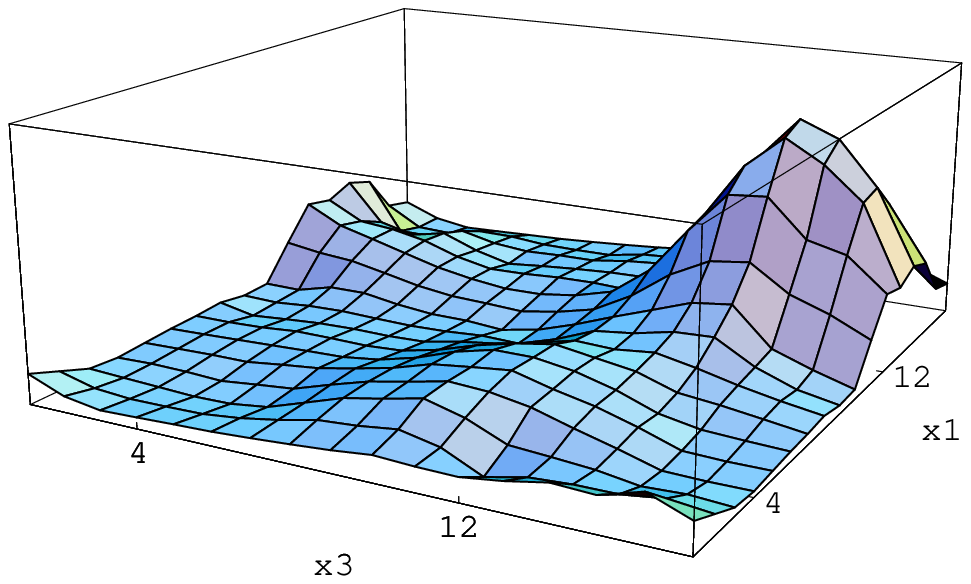}
\includegraphics[width=0.32\linewidth]
{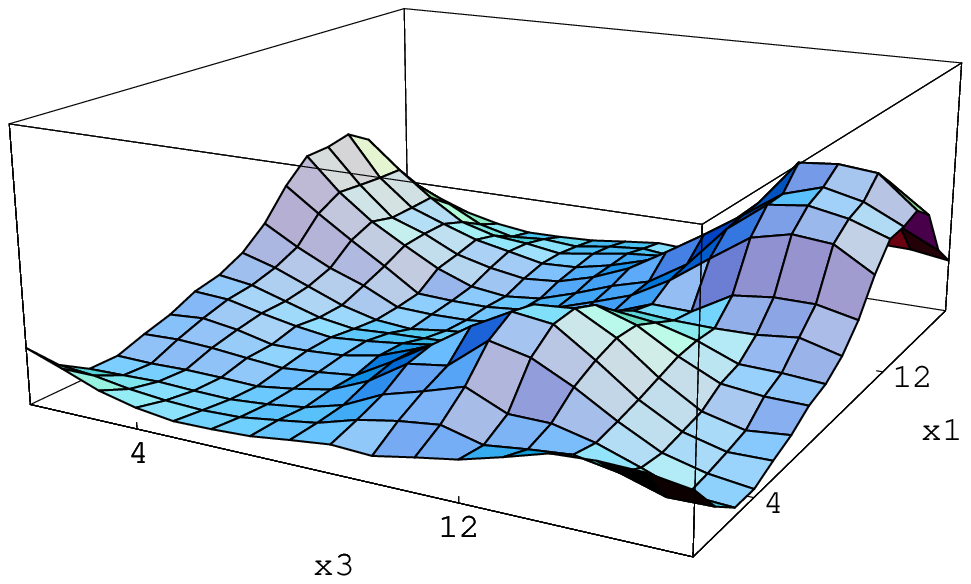}\\
\includegraphics[width=0.32\linewidth]
{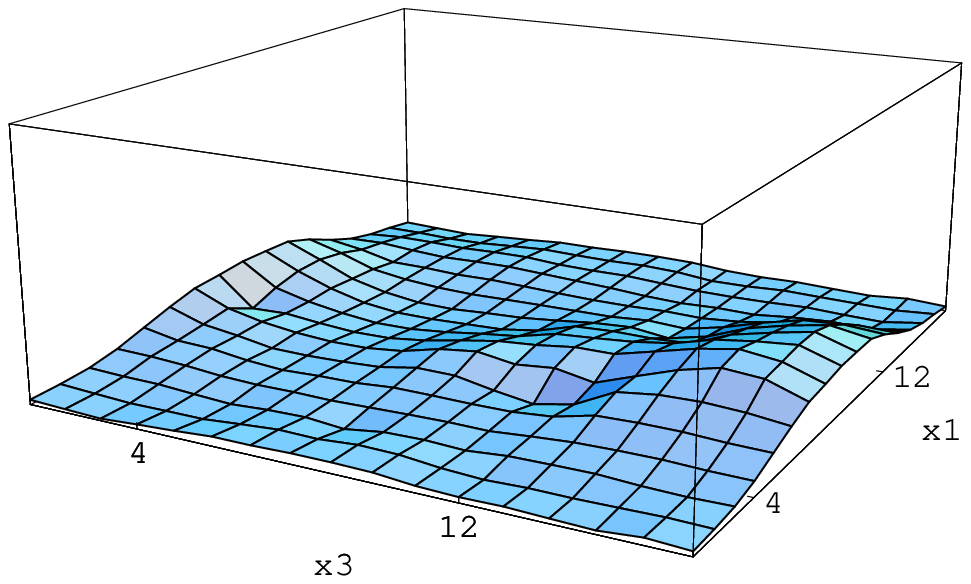}
\includegraphics[width=0.32\linewidth]
{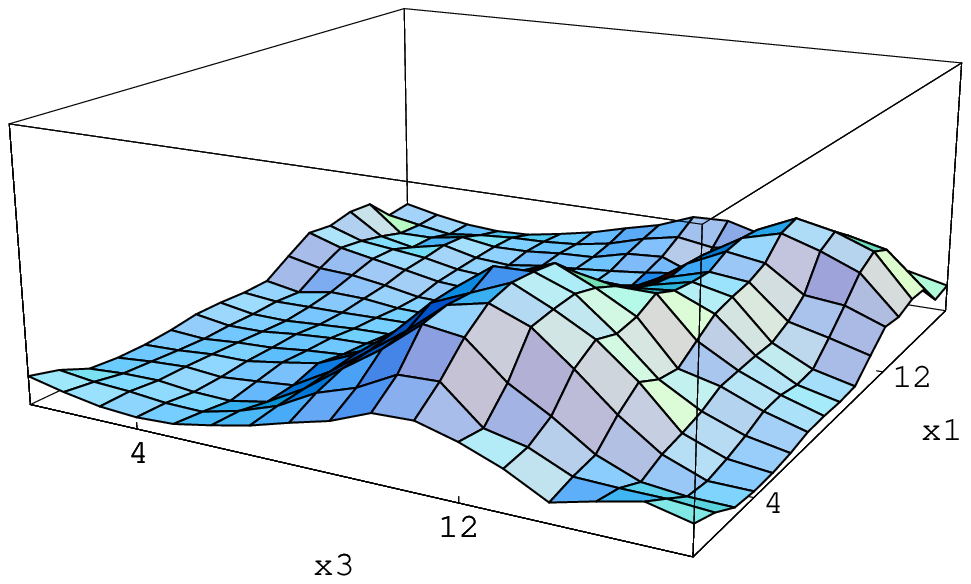}
\includegraphics[width=0.32\linewidth]
{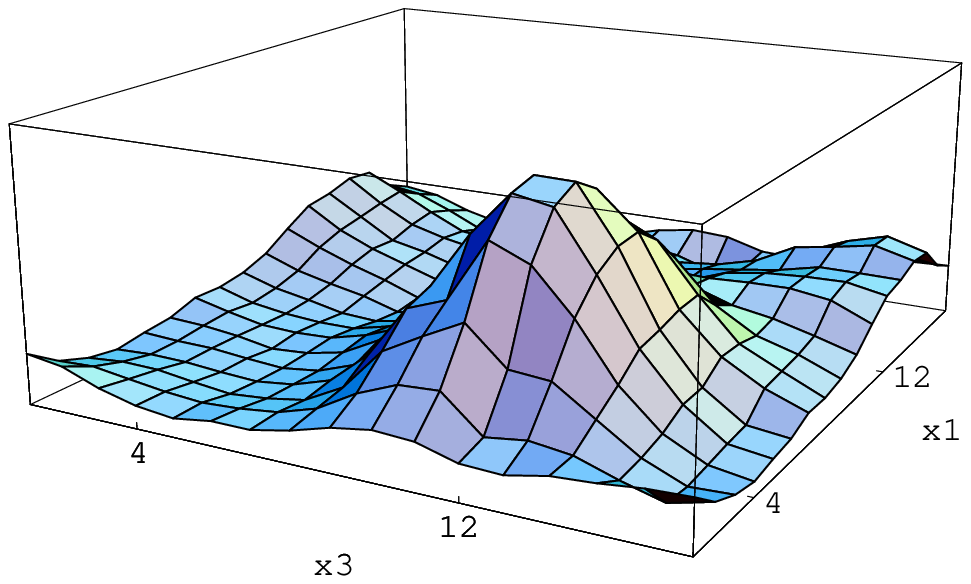}
\caption{The lowest Laplacian mode in a thermalized background with
$\zeta=0.02,0.355,0.48$ from left to right and in
  lattice planes $(x_2,x_4)=(12,1),(4,4),(6,3)$ from top to
  bottom. The vertical scale is 0.05.}
\label{fig_2}
\end{figure}

In this subsection we explore the Laplacian modes as an analyzing  tool on
thermalized gauge field configurations. Fig.\ \ref{fig_2} shows the
lowest-lying mode in a generic background obtained on a $16^3\cdot 4$
lattice at $\beta=2.2$ (which amounts to $T=0.75\, T_c$) confirming
the expectation that these modes are free of
UV fluctuations.
We allow for a complex phase in the boundary condition,
$\phi(x_4+N_4)=\exp(2\pi i\zeta) \phi(x_4)$ parametrised by an angle
$\zeta\in[0,1/2]$ (the remaining $\zeta$'s can be mapped to this
interval by charge conjugation). The `evolution' of the Laplacian mode
with this angle is shown vertically in Fig.\ \ref{fig_2}, while the 3 rows
represent different but fixed lattice planes, which contain the global
maximum of the mode. In the example we find 3 lattice locations which
become the global maximum within some $\zeta$-interval. Therefore, the
lowest Laplacian mode in a thermalized background is hopping, too. 

At the hopping  points in $\zeta$ the value of the
maximum\footnote{
The global minimum is not stable enough to employ it for analyzing purposes.}
as well as the inverse participation ratio 
(taking values between ${\rm IPR}=4$ and ${\rm IPR}=11$) 
decrease and the mode rearranges itself.
For some boundary conditions the lowest Laplacian mode can be
characterized as a global structure.

Analyzing 50 independent configurations the lowest mode 
hops up to four times as a function of $\zeta$. 
Apart from the
Laplacian mode being wave-like, the hopping phenomenon resembles
the behaviour of the fermion zero mode \cite{gattringer:02b}. However,
in both cases it is not obvious which gluonic feature, say in the
topological charge, discriminates the preferred
locations. Interestingly, we have
found a correlation of the maximum of the periodic and antiperiodic
Laplacian mode to positive and negative Polyakov loops,
respectively. This is the same tendency as for
calorons (see Fig. \ref{fig_1}) which could be understood as an
argument for calorons
`underlying' the quantum configuration. 
Alternatively, the Polyakov
loop could play a role defining pinning centers for the Laplacian
mode in the spirit of Anderson localisation in a random
potential \cite{anderson:58}.
  
The lowest Laplacian mode in the {\em adjoint} representation we have
found to have {\em minima} at the caloron monopoles (see also
\cite{bruckmann:01a,alexandrou:00a}) 
that extend to two-dimensional sheets for antiperiodic
boundary conditions. In thermalized backgrounds, the maxima of the
adjoint modes are
correlated to the fundamental ones with the same boundary condition,
but are stronger localized than the latter.

\section{A Fourier-like filter}

Laplacian (`harmonic') eigenmodes can be used to 
define a new Fourier-like filter
which we present now \cite{bruckmann:05b}. 
We were inspired by the representation of the
field strength in terms of fermionic modes in \cite{gattringer:02c}, 
but here we aim to reconstruct directly the link variables on which
any observable can be measured. To this
end we combine the definition of the lattice Laplacian, Eq.\ (\ref{eqn_1}), with a
spectral decomposition and at $y=x+\hat{\mu}$ immediately obtain 
\begin{eqnarray}
U_\mu^{ab}(x)=
-\sum_{n=1}^\mathcal{N}\lambda_n\phi^a_n(x)\phi^{*b}_n(x+\hat{\mu})\,,
\qquad \mathcal{N}=N_1N_2N_3N_4\cdot 2
\label{eqn_2} 
\end{eqnarray}
The idea is to truncate the sum on the right hand side 
of this equation at some $N\ll\cal{N}$.
The question arises, how to relate such an expression to a
{\em unitary} link variable. 
Here the charge conjugation helps, since it guarantees that
every eigenvalue is two-fold degenerate with eigenfunction related as
$\phi'^a(x)=\epsilon^{ab}\phi^{*b}(x)$. It follows that the corresponding
bilinear expressions in Eq.\ (\ref{eqn_2}) add up to an element of
$SU(2)$ up to a factor. The same applies to the staple average in cooling
or smearing.
We divide by the square root of this factor, which is the determinant and
positive for all practical purposes, and obtain the
final filter formula
\begin{eqnarray}
\tilde{U}_\mu^{ab}(x)_N=
\left(-\sum_{n=1}^N\lambda_n\left[\phi^a_n(x)\phi^{*b}_n(x+\hat{\mu})-
\epsilon^{ac}\phi^c_n(x)\phi^{*d}_n(x+\hat{\mu})\epsilon^{db}\,\right]
\right)/\sqrt{\det (\ldots)}
\label{eqn_3} 
\end{eqnarray}
The quality of the filter is controlled by $N$, where $N=\cal{N}$
reproduces the original configuration exactly. 
 In the other extreme case, $N=1$, the filtered
links $\tilde{U}_\mu(x)$ can be shown to be pure gauge.
From the behaviour of $\phi^a(x)$ under gauge transformations it
follows immediately that $\tilde{U}_\mu(x)$ transforms as a
link (and no gauge fixing is involved in the filter).

The filtering can be performed with Laplacian modes of any boundary
condition. Then Eq.\ (\ref{eqn_3}) is just the average over opposite
boundary conditions $\zeta$ and $-\zeta$. This additional parameter
completely fixes the Polyakov loop at $N=1$ to 
${\rm tr}\,\tilde{\mathcal{P}}(\vec{x})/2=\cos(2\pi\zeta)$, 
while for nontrivial cases $N\geq 2$ 
the Polyakov loop is observed to fluctuate around this value. It
follows that in order to optimize the filter for our circumstances
(confined phase, nontrivial holonomy calorons) the best choice is
$\zeta=1/4$, 
i.e.\ halfway between periodic and antiperiodic boundary conditions. 

\begin{figure}[t]
\begin{minipage}[t]{0.45\linewidth}
\includegraphics[width=1\linewidth]
{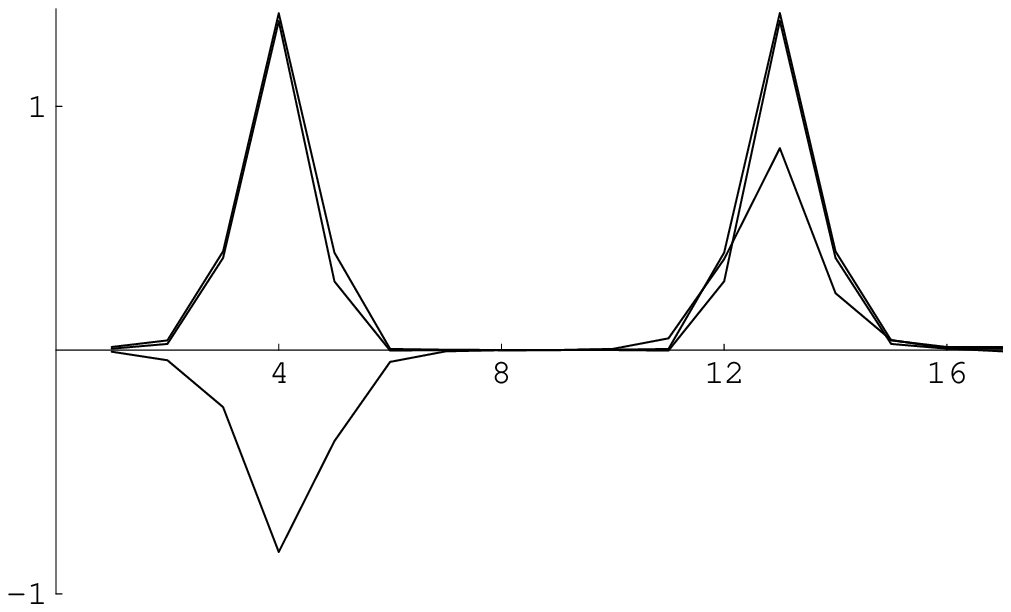}
\end{minipage}\hfill
\begin{minipage}[t]{0.45\linewidth}
\includegraphics[width=1\linewidth]
{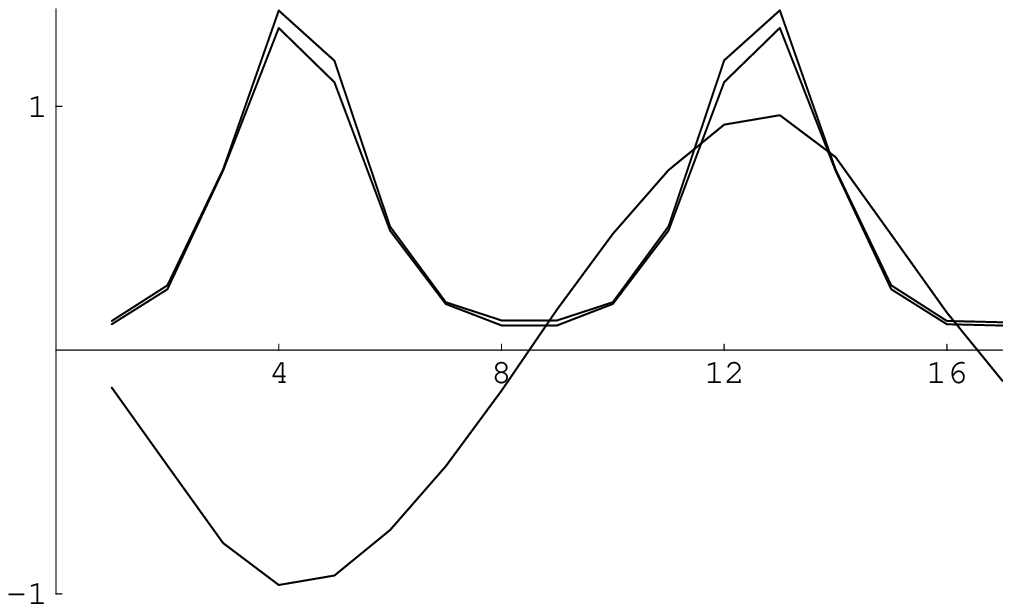}
\end{minipage}
\caption{Action and topological density and Polyakov loop of the
  caloron of Fig.\ \protect\ref{fig_1} (left) 
filtered with $N=4$
  (left, densities multiplied by 100) and $N=150$ (right, densities
  multiplied by 400) modes ($\zeta=1/4$).}
\label{fig_3}
\end{figure}

In order to test the filter we again start with the caloron
background. Fig.\ \ref{fig_3} shows the action and topological
density as well as the Polyakov loop measured on the filtered links
$\tilde{U}_\mu(x)$, to be compared with Fig.\
\ref{fig_1} (left). The filter with the number of modes as low as $N=4$
starts to reproduce the classical structures qualitatively, while for
$N=150$ the agreement is almost perfect.

\begin{figure}[b]
\centering
\includegraphics[width=0.7\linewidth]
  {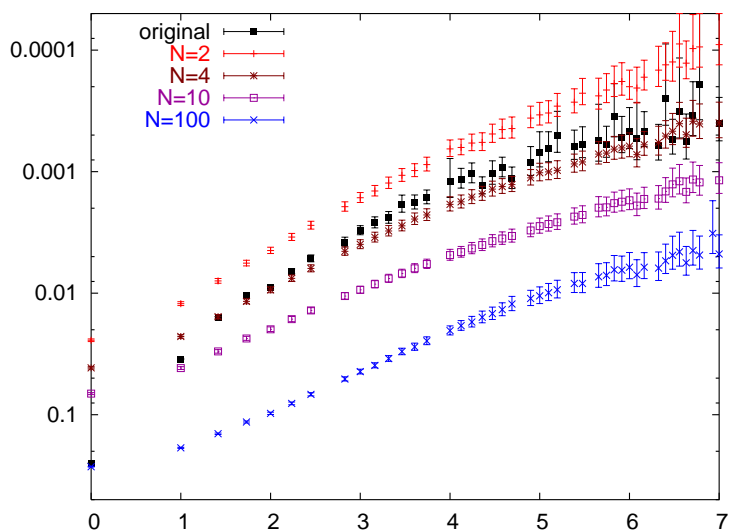}
\caption{Polyakov loop correlator plotted on an inverse logarithmic
  scale over the distance for different $N$ (at $\zeta=1/4$) compared
  to the original one.}
\label{fig_4}
\end{figure}

What is even more remarkable is that the filter method preserves the string
tension. We plot the logarithm of the Polyakov loop correlator
measured on 50 configurations at $T=0.75\,T_c$ in Fig.\ \ref{fig_4}.
It reveals a clearly linear behaviour with a slope
reproducing the original one within 15\%. 
The minimal number of modes, $N=2$, suffices for this behaviour.
We conclude that the
confining properties of lattice gauge theory are captured by the
lowest Laplacian modes.
The Polyakov loop correlator after filtering has no sign of a Coulomb
regime, since the filter has washed out short range fluctuations. At
zero distance the values 
$\langle({\rm tr}\,\tilde{\mathcal{P}}(\vec{x}))^2\rangle$
are smaller than the original one which
indicates that the distribution of Polyakov loops over the lattice
sites has become narrower.
 
Since the filter is not biased to classical solutions nor to
particular degrees of freedom in the gauge field 
(like monopoles or vortices), it is
interesting to have a closer look at the vacuum structures that emerge when
the filter is applied to generic configurations. In both the Polyakov
loop and the action density, more and more fluctuations appear with an
increasing number of modes kept in the filter. 
It might well be that the fluctuation
at $N=2$ are the minimally required ones for the long-range physics.

Concerning the action density, these objects are isolated peaks (which
are non-static and not necessarily (anti)selfdual). This
phenomenon is counterintuitive as the filter uses the lowest-lying
Laplacian modes, however, this is also the case for the classical
calorons, see Fig.\ \ref{fig_3} (left).
We have found a correlation of the peak structure of the
filtered action density to that
emerging after cooling or smearing in an early stage.
More work has to be done to better understand the ($N$-dependent)
spiky structures induced by the filter.


\providecommand{\href}[2]{#2}\begingroup\raggedright\endgroup

\end{document}